\documentclass[12pt]{elsarticle}
\usepackage{refmerge}
\usepackage{graphicx}
\usepackage{amssymb,amsmath}

\begin{document}

\title{\bf\boldmath Search for direct production
of the $f_1(1285)$ resonance in $e^+e^-$ collisions}

\begin{abstract}
A search for direct production of the $f_1(1285)$
resonance in $e^+e^-$ annihilation is performed with the SND
detector at the VEPP-2000 $e^+e^-$ collider. The analysis is 
based on data with an integrated luminosity of 15.1 pb$^{-1}$ accumulated in 
the center-of-mass energy range 1.2--1.4~GeV. 
Two $e^+e^-\to f_1(1285)$ candidate events are found at the peak of the 
resonance and zero events beyond the resonance. The significance of
the $e^+e^-\to f_1(1285)$ signal is $2.5\sigma$.
The cross section at the maximum of the resonance is found to be
$\sigma(e^+e^-\to f_1)=45^{+33}_{-24}$~pb. The corresponding branching
fraction $B(f_1(1285)\to e^+e^-)=(5.1^{+3.7}_{-2.7})\times 10^{-9}$.
We consider this result as a first indication of the process
$e^+e^-\to f_1(1285)$. The measured branching fraction is consistent 
with the theoretical prediction.
\end{abstract}

\author[adr1,adr2]{M.~N.~Achasov}
\author[adr1,adr2]{A.~Yu.~Barnyakov}
\author[adr1,adr2]{K.~I.~Beloborodov}
\author[adr1,adr2]{A.~V.~Berdyugin}
\author[adr1,adr2]{D.~E.~Berkaev}
\author[adr1]{A.~G.~Bogdanchikov}
\author[adr1]{A.~A.~Botov}
\author[adr1,adr2]{T.~V.~Dimova}
\author[adr1,adr2]{V.~P.~Druzhinin}
\author[adr1,adr2]{V.~B.~Golubev}
\author[adr1,adr2]{L.~V.~Kardapoltsev}
\author[adr1]{A.~S.~Kasaev}
\author[adr1,adr2]{A.~G.~Kharlamov}
\author[adr1,adr2,adr3]{I.~A.~Koop}
\author[adr1,adr2]{A.~A.~Korol}
\author[adr1]{D.~P.~Kovrizhin}
\author[adr1]{S.~V.~Koshuba}
\author[adr1,adr2]{A.~S.~Kupich}
\author[adr1]{R.~A.~Litvinov}
\author[adr1]{ A.~P.~Lysenko}
\author[adr1]{K.~A.~Martin}
\author[adr1,adr2]{N.~A.~Melnikova}
\author[adr1,adr2]{N. Yu. Muchnoi}
\author[adr1]{A.~E.~Obrazovsky}
\author[adr1]{E.~V.~Pakhtusova}
\author[adr1,adr2]{E.~A.~Perevedentsev}
\author[adr1,adr2]{K.~V.~Pugachev}
\author[adr1,adr2]{S.~I.~Serednyakov}
\author[adr1,adr2]{Z.~K.~Silagadze}
\author[adr1,adr2]{P.~Yu.~Shatunov}
\author[adr1,adr2]{Yu.~M.~Shatunov}
\author[adr1,adr2]{D.~A.~Shtol}
\author[adr1,adr2]{D.~B.~Shwartz}
\author[adr1,adr2]{I.~K.~Surin}
\author[adr1,adr2]{ Yu.~V.~Usov}
\author[adr1,adr2]{I. M. Zemlyansky}
\author[adr1,adr2]{V.~N.~Zhabin}
\author[adr1,adr2]{V. V. Zhulanov}
\address[adr1]{Budker Institute of Nuclear Physics, SB RAS,
Novosibirsk, 630090, Russia}
\address[adr2]{Novosibirsk State University, Novosibirsk, 630090, Russia}
\address[adr3]{Novosibirsk State Technical University,
Novosibirsk, 630092, Russia}
\fntext[tnot]{Corresponding author: druzhinin@inp.nsk.su}
\maketitle

\section{Introduction}
The dominant mechanism of hadron production in $e^+e^-$ collisions is
single-photon annihilation. Annihilation through two photons is 
suppressed by a factor of $\alpha^2$, where $\alpha$ is the fine structure 
constant. The only observed process of the two-photon annihilation into 
hadrons in $e^+e^-$ collisions is the production of two vector mesons, 
$\rho^0\rho^0$ and $\rho^0\phi$, in the BABAR experiment~\cite{babar}. 
Experiments on the search for production of a single $C$-even resonance 
began more than 30 years ago at the VEPP-2M $e^+e^-$ collider with the ND 
detector\cite{ND}. In these experiments, the first upper limits were set on 
the probabilities of the inverse reactions, namely the decays $\eta'$, 
$f_0(975)$, $f_2(1270)$, $f_0(1300)$, 
$a_0(980)$, and $a_2(1320)$ to $e^+e^-$ pairs. In recent experiments at 
the colliders VEPP-2M~\cite{ND,SND1}, VEPP-2000~\cite{SND2,SND3}
and BEPCII~\cite{bes}, this inverse-reaction technique was used to set 
the best upper limits on the electron widths of the resonances listed above,
as well as $\eta$ and $X(3872)$. 
\begin{figure}[htbp]
\centering
\includegraphics[width=0.65\textwidth]{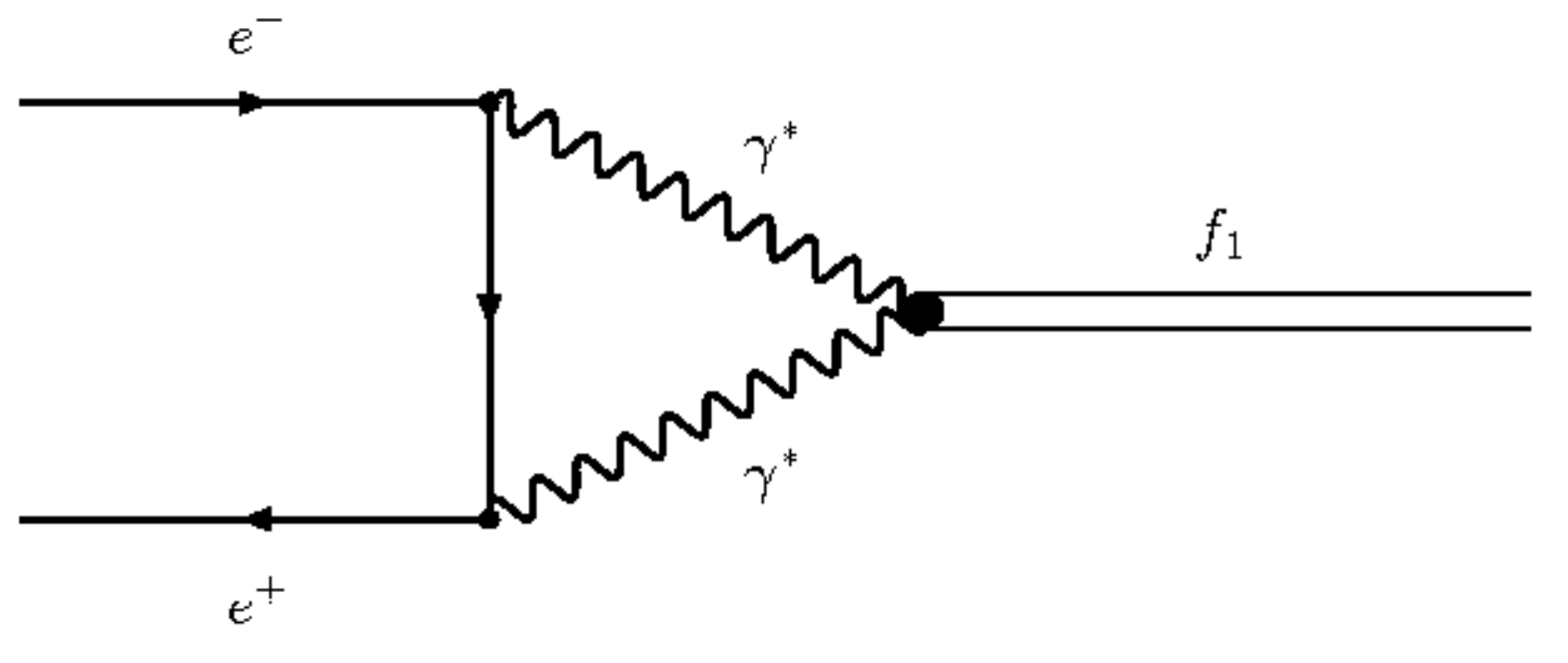}
\caption{ \label{fig1}
The diagram for the process $e^+e^-\to f_1$.}
\end{figure}

In this paper, we present the first search for the process 
$e^+e^-\to f_1(1285)$ shown in the diagram in Fig.~\ref{fig1}. 
There are also no data on the $f_1(1285)\to e^+e^-$ decay. 
Axial-vector resonances, like the $f_1(1285)$, do not decay
into two photons, but may nevertheless be produced in $e^+e^-$ annihilation,
through virtual photons. Theoretically, the $f_1(1285)\to e^+e^-$ decay,
as well as the $e^+e^-\to f_1(1285)$ process, are discussed in 
Ref.~\cite{rudenko1,rudenko2} within the framework of the 
vector-meson-dominance model
assuming that the virtual photons in Fig.~\ref{fig1} are coupled with the 
$f_1$ meson via intermediate $\rho^0$ mesons. The two coupling constants 
describing the $\gamma^\ast\gamma^\ast\to f_1(1285)$ amplitude
and the relative phase between them are determined from the experimental data 
on the $f_1(1285)\to\rho\gamma$
decay~\cite{ves,pdg} and the $\gamma\gamma^\ast\to f_1(1285)$ 
reaction~\cite{l3}. The $e^+e^-\to f_1(1285)$ cross section is predicted to
be $(31\pm 16)$~pb. The corresponding $f_1(1285)\to e^+e^-$ branching fraction 
calculated using the formula 
$\sigma(e^+e^-\to f_1)=(12\pi/m^2_{f_1})B(f_1\to e^+e^-)$  is
$(3.5\pm 1.8)\times 10^{-9}$. The $f_1$-meson mass and width are 
$1281.9\pm0.5$~MeV and $22.7\pm1.1$~MeV~\cite{pdg}, respectively.
Its main decay modes are $\pi^+\pi^-\pi^0\pi^0$,
$\pi^+\pi^-\pi^+\pi^-$, $\eta\pi^+\pi^-$, and $\eta\pi^0\pi^0$~\cite{pdg}.
The first three of these final states are also produced in the single-photon 
annihilation and have cross sections at $\sqrt{s}=m_{f_1}$ by two to three 
orders of magnitude higher than the prediction for $\sigma(e^+e^-\to f_1)$.
Therefore, the most viable mode for searching for the process $e^+e^-\to f_1(1285)$ 
is $f_1\to\eta\pi^0\pi^0$ with the branching fraction $(17.3\pm 0.7)\%$.

The search for the direct production of the $f_1(1285)$  meson in 
$e^+e^-$ collisions is performed at the SND experiment at 
the VEPP-2000 collider~\cite{VEPP2000}.

\section{Detector and experiment}
We analyze data with an integral luminosity of 15.1~pb$^{-1}$, recorded in 
2010, 2011, 2012, and 2017 in the center-of-mass energy region 
$\sqrt{s} =$1.2--1.4~GeV at 12 energy points listed in Table~\ref{tab2}. The 
data set at the maximum of the $f_1$ resonance $\sqrt{s} =1.282$~GeV with an 
integral luminosity of 3.4~pb$^{-1}$ was collected in 2017 specifically for 
this analysis. For 2010-2012 energy scans the collider energy was controlled 
with an accuracy of 2--4 MeV. In 2017 the beam energy was monitored with an 
accuracy of about 80 keV by the backscattering-laser-light system~\cite{laser}.
The center-of-mass energy spread at $\sqrt{s} =1.282$~GeV is 630~keV.

A detailed description of the SND detector can be found in Refs.~\cite{SND}.
This is a nonmagnetic detector, the main part of which is a three-layer 
spherical electromagnetic calorimeter based on NaI (Tl) crystals. 
The calorimeter covers about 95\% of the solid angle and has an energy 
resolution for photons of $\sigma_E/E=4.2\%/\sqrt[4]{E({\rm GeV})}$,
and an angular resolution of $1.5^\circ$. The directions of charged particles
are measured in a tracking system consisting of a nine-layer drift chamber and 
a proportional chamber with cathode-strip readout. The solid angle covered by
the tracking system is 94\% of $4\pi $. The calorimeter is surrounded by 
a muon system.

The search for the process $e^+e^-\to f_1(1285)$ is performed 
in the channel $f_1(1285)\to\eta\pi^0\pi^0$ with the subsequent decays 
$\eta\to\gamma\gamma$ and $\pi^0\to\gamma\gamma$. Since the final state for 
the process under study does not contain charged particles, the
process without charged particles $e^+e^-\to \gamma\gamma$ is used for 
normalization. As a result of this normalization, systematic uncertainties
associated with the event selection in the hardware trigger and beam-generated 
spurious charged tracks cancel in the $e^+e^-\to f_1$ cross section.
The systematic uncertainty of the luminosity measurement is studied in
detail in Ref.~\cite{SNDomegapi} and is equal to 2.2\%.
The distribution of the integrated
luminosity over 12 energy points is given in Table~\ref{tab2}. 
About 30\% of the analyzed data sample is collected near the $f_1(1285)$
maximum at $\sqrt{s} = 1.280$ and 1.282 GeV.

According to the Particle Data Group (PDG) table~\cite{pdg}, 
the dominant intermediate state in the $f_1(1285)\to\eta\pi\pi$ decay
is $a_0\pi$. Its fraction is $(73\pm8)\%$. The process 
$e^+e^- \to f_1(1285) \to a_0^0\pi^0 \to \eta\pi^0\pi^0$ is simulated using a 
Monte-Carlo (MC) event generator based on the formulas from 
Ref.~\cite{rudenko1}. It takes into account the identity of the final 
pions and the finite $a_0$ width. The $a_0$ line shape is described by a 
Breit-Wigner function.
To simulate the remaining 27\% of the $f_1(1285)\to\eta\pi\pi$ 
decay, a MC generator with the $f_0(500)\eta$ intermediate state is used.

Event generators for signal and background processes
include radiative corrections~\cite{rad1}, in particular, the
emission of an additional photon from the initial state~\cite{rad2}.
The Born cross sections needed for simulation of background processes
are taken from existing data. For main background processes 
$e^+e^-\to\omega\pi^0\to\pi^0\pi^0\gamma$, 
$e^+e^-\to\omega\pi^0\pi^0\to 3\pi^0\gamma$,
$e^+e^-\to\eta\gamma\to3\pi^0\gamma$ the recent most accurate cross section
measurements~\cite{SNDomegapi,ompipi1,ompipi2,etag1,etag2} are used.
Other processes with multiphoton final states, e.g.
$e^+e^-\to\omega\eta$, $e^+e^-\to\phi\eta$, and $e^+e^-\to\omega\eta\pi^0$,
have small cross sections in the energy range of interest and
give negligible contributions to the background. The processes with
$K_L$ mesons, $e^+e^-\to K_S K_L$, $e^+e^-\to K_S K_L\pi^0$, etc, are
strongly suppressed by our selection criteria (see Sec.~\ref{sel})
and are also negligible.

The signal process is studied in the six-photon final state 
$f_1(1285)\to \eta\pi^0\pi^0 \to 6\gamma$. In the dominant background 
process $e^+e^-\to\omega\pi^0\to \pi^0\pi^0\gamma$ with five photons in the 
final state, an additional photon appears either because 
of initial state radiation, or because of splitting of electromagnetic showers,
or because of superimposing beam-generated background. To simulate the latter
effect, special background events are used, which were recorded during 
experiment with a random trigger. These events are superimposed on the 
simulated events. 

\section{Event selection\label{sel}}
To search for the process $e^+e^- \to f_1(1285) \to \eta\pi^0\pi^0$,
events with exactly six reconstructed photons and no tracks in the tracking
system are selected. Photons are clusters in the calorimeter with 
the energy deposition greater than 20 MeV. The
total energy deposition in the calorimeter $E_ {tot}$ and the total 
event momentum $P_ {tot}$ calculated using the energy depositions 
in the calorimeter crystals must satisfy the conditions
\begin{equation}
0.7 < E_{tot}/\sqrt{s} < 1.2,~P_{tot}/\sqrt{s} < 0.3,~
(E_{tot} - P_{tot})/\sqrt{s} > 0.7,
\label{eton_vs_ptrt}
\end{equation}
which provide approximate energy and momentum balance in an event.
To suppress cosmic-ray background, no signal in the muon system is
required. 

\begin{figure}
\centering
\includegraphics[width=0.47\textwidth]{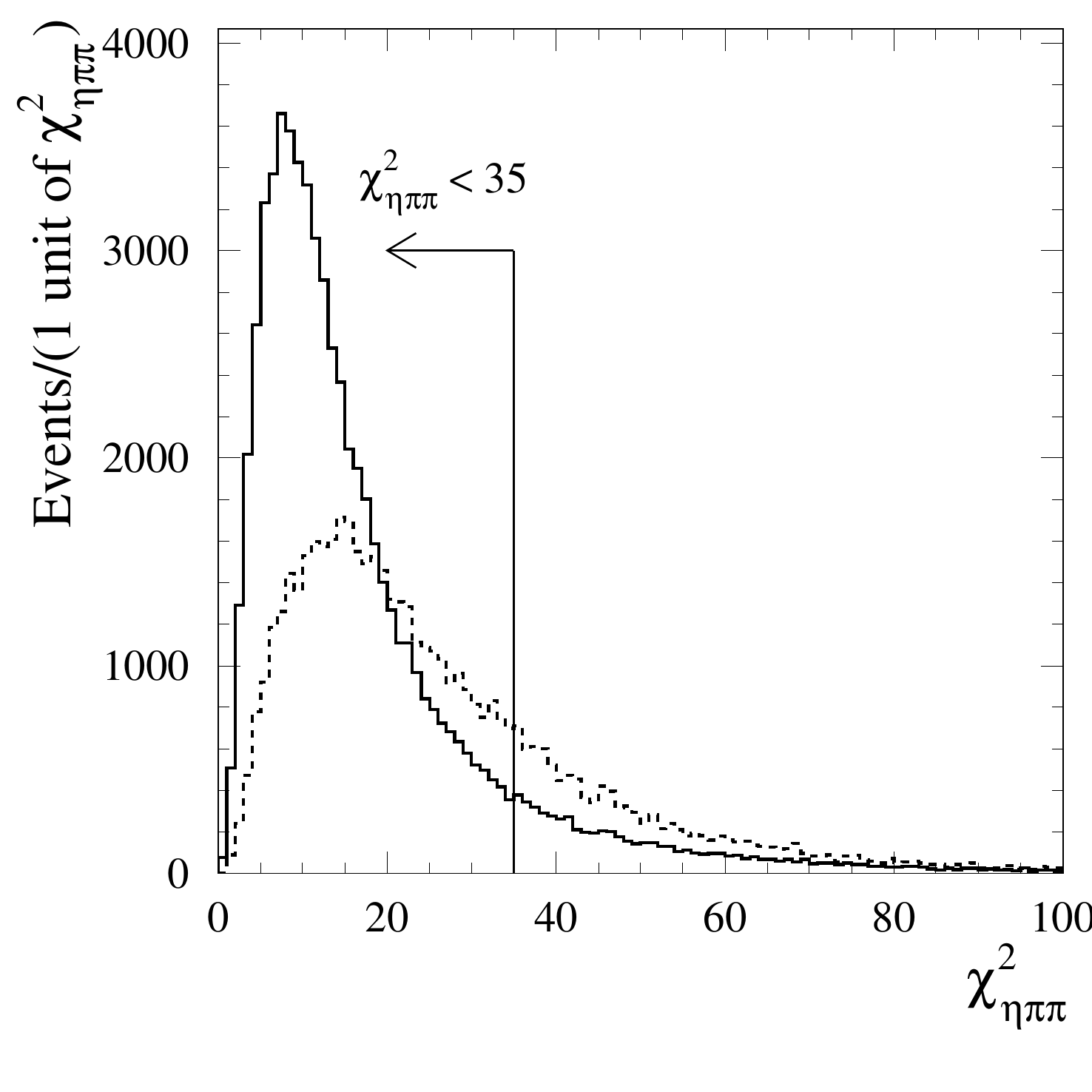}
\includegraphics[width=0.47\textwidth]{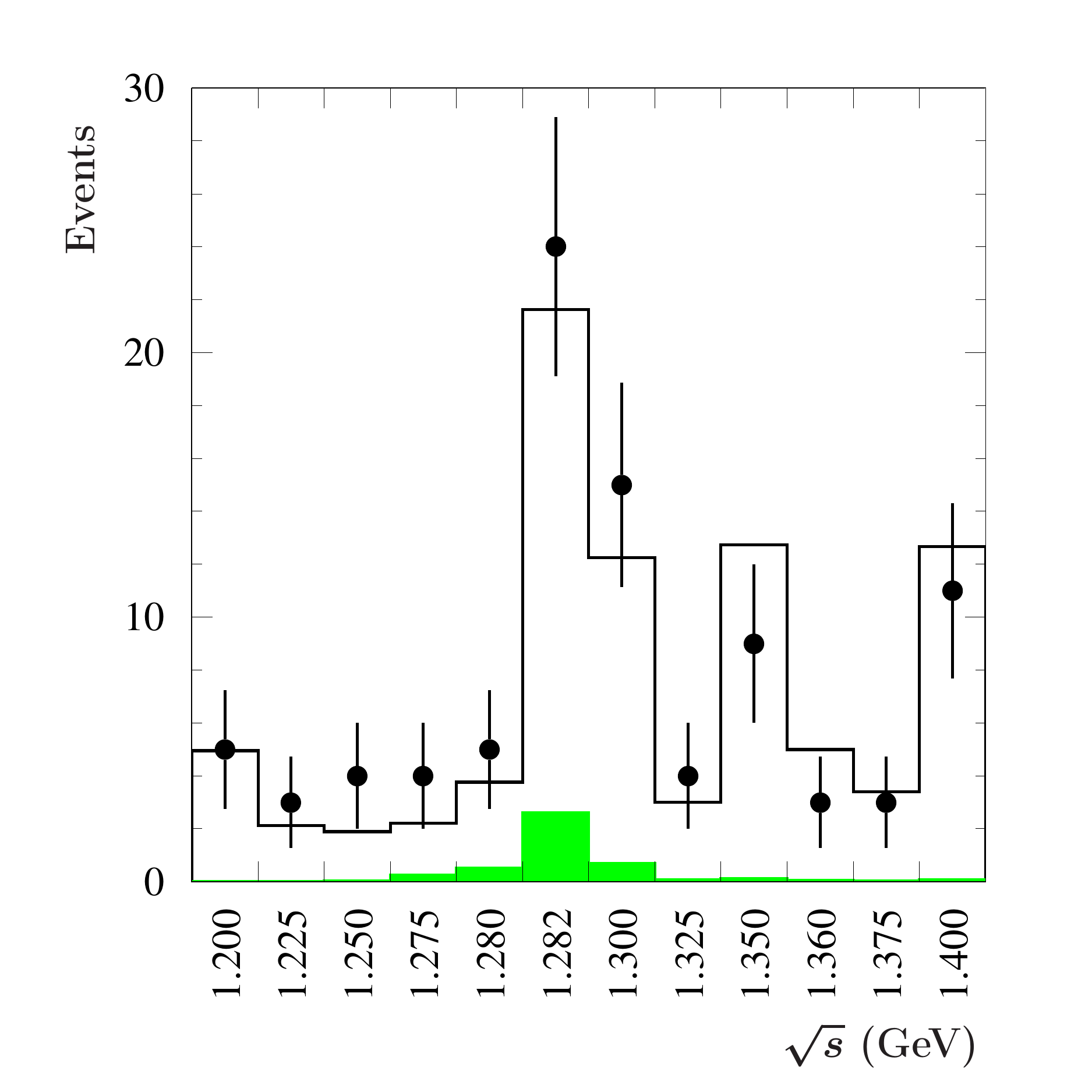}
\caption{Left panel: The $\chi^2_{\eta\pi\pi}$ distribution for simulated 
signal $e^+e^-\to\eta\pi^0\pi^0$ events (solid histogram)and
background  $e^+e^-\to \omega\pi^0(\gamma)$ events (dashed histogram).
The arrow indicates the selection criterion used.  Right panel: 
The distribution of 90 data events selected by the condition 
$\chi^2_{\eta\pi\pi}<35$ over the 12 energy points (points with error bars).
The open histogram is the expected distribution for background events. 
The shaded histogram represents the expected signal distribution for
$\sigma(e^+e^-\to f_1)=50$ pb.
\label{fig2}}
\end{figure}

Then we select events containing two $\pi^0$ candidate and one $\eta$
candidate, which are defined as two-photon pairs with invariant masses in the
windows $|M_{2\gamma}-m_{\pi^0}|<35$ MeV and $|M_{2\gamma}-m_\eta|<50$ MeV,
respectively. For the selected events, a kinematic fit to the hypothesis
$e^+e^-\to\eta\pi^0\pi^0$ is performed with a requirement of total energy
and momentum conservation and three invariant-mass constraints.
The condition on the $\chi^2$ of the kinematic fit, $\chi^2_{\eta\pi\pi}<35$,
is applied. The distribution of this parameter for the simulated signal events
and background events from the process $e^+e^-\to\omega\pi^0$ 
are shown in Fig.~\ref{fig2}~(left). It should be noted that the background 
is suppressed by the condition that six photons form two $\pi^0$ and one
$\eta$ candidates.  Background events satisfying this condition have
$\chi^2_{\eta\pi\pi}$ distribution not strongly different from the
distribution for signal events.
The distribution of 90 selected data events over the 12 energy points is 
shown in Fig.~\ref{fig2} (right).

Background events passing the selection criteria come from the processes
$e^+e^-\to\omega\pi^0$, $e^+e^-\to\omega\pi^0\pi^0$, and $e^+e^-\to\eta\gamma$.
The number of background events estimated from simulation is $86\pm 1$,
about 90\% of which are from the process $e^+e^-\to\omega\pi^0$. 
The expected number of signal events for $\sigma(e^+e^-\to f_1)=50$ pb is 4.7. 
The calculated distributions of background and signal events over the 12 
energy points are shown in Fig.~\ref{fig2} (right). It is seen that the 
data and simulated background distributions are in good agreement.
At this stage of the selection, the background is too large to observe the 
signal of the $f_1(1285)$ decay.  
\begin{figure}
\centering
\includegraphics[width=0.47\textwidth]{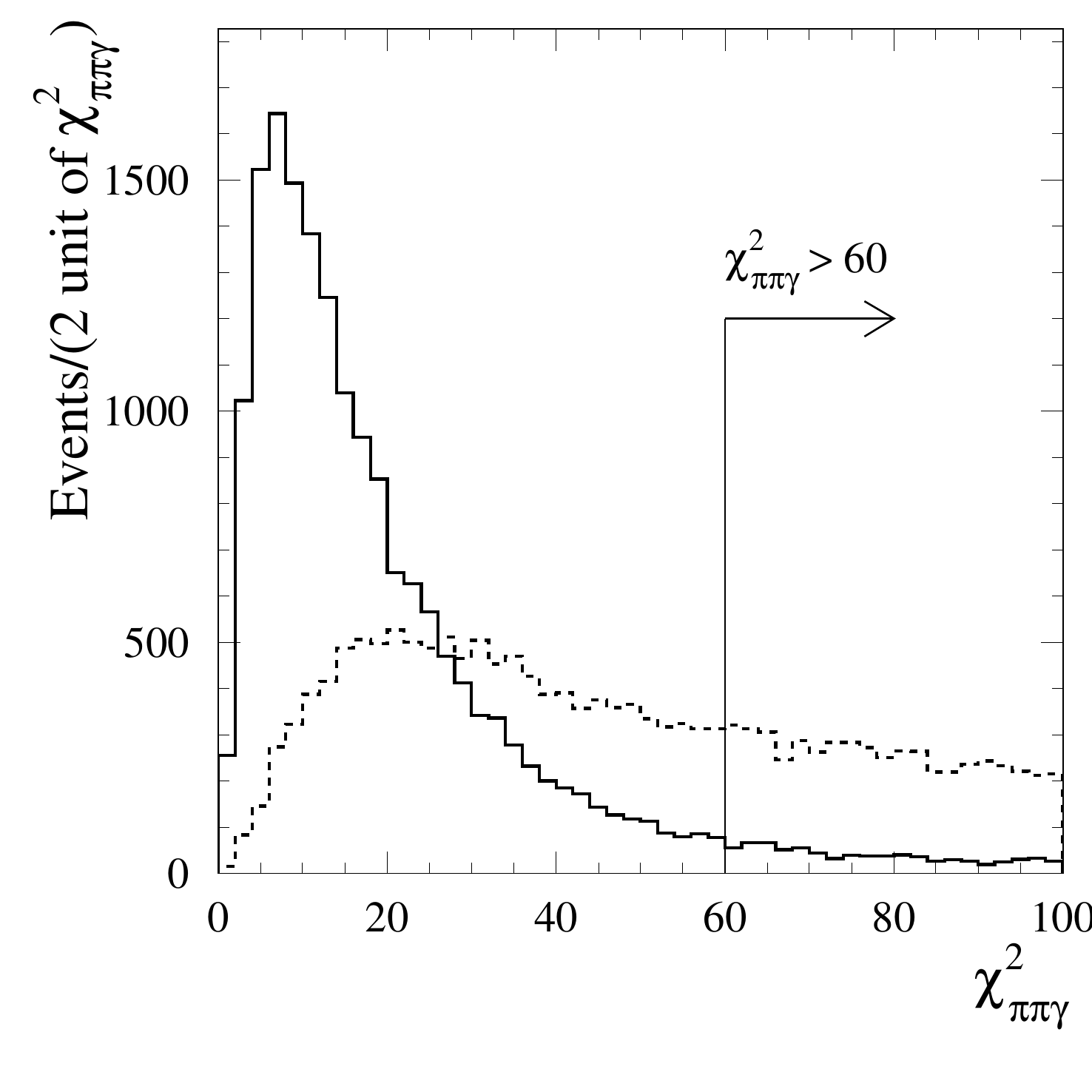}
\includegraphics[width=0.47\textwidth]{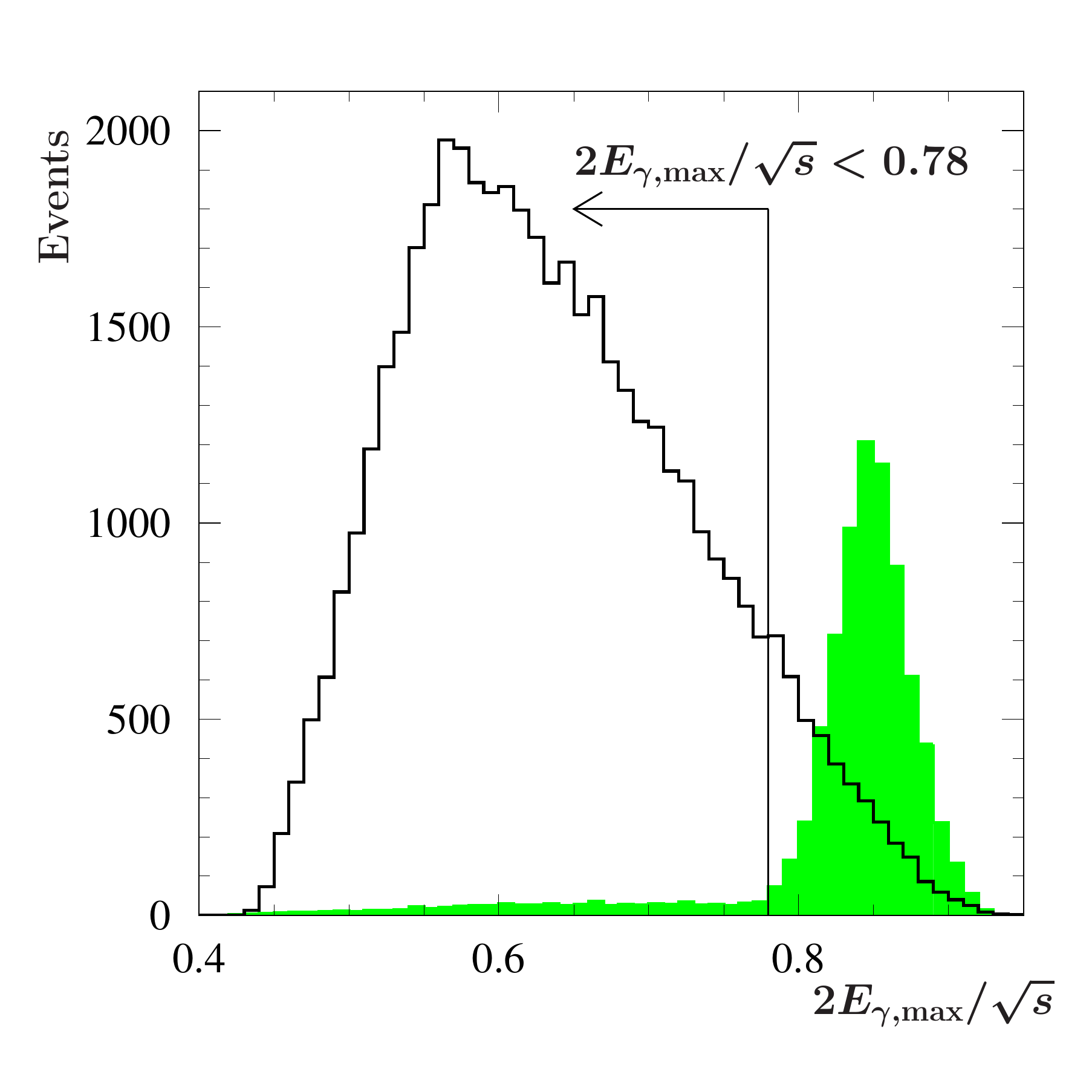}
\caption{Left panel: The distribution of the parameter $\chi^2_{\pi\pi\gamma}$
for simulated events of the processes $e^+e^-\to \omega\pi^0(\gamma)$ (solid 
histogram) and $e^+e^-\to f_1 \to \eta\pi^0\pi^0$ (dashed histogram). 
The arrow indicates the selection criterion used.
Right panel: The distribution of the normalized energy of the most energetic 
photon in an event for simulated events of the processes 
$e^+e^- \to \eta\pi^0\pi^0$ (open histogram) and 
$e^+e^-\to \eta\gamma(\gamma)$ (shaded histogram). 
The arrow indicates the selection criterion used.
\label{fig3}}
\end{figure}

Since the main background comes from the process 
$e^+e^-\to\omega\pi^0\to\pi^0\pi^0\gamma$,
the kinematic fit in the hypothesis $e^+e^-\to\pi^0\pi^0\gamma$ is also
performed. During the fit, all possible five-photon combinations are tested.
Events containing a combination with $\chi^2_{\pi\pi\gamma}<60$
are rejected. The $\chi^2_{\pi\pi\gamma}$ distributions 
for simulated signal and background $e^+e^-\to\omega\pi^0$ events 
are shown in Fig.~\ref{fig3} (left). 

To calculate other two parameters helpful for background suppression, we use 
the energies and angles of photons after the kinematic fit to the 
$e^+e^-\to 6\gamma$ hypothesis. Figure~\ref{fig3} (right) shows the 
distribution of the normalized energy of the most energetic photon in an event 
$2E_{\gamma,\rm{max}}/\sqrt{s}$ for simulated signal and background
$e^+e^-\to \eta\gamma(\gamma)$ events. To suppress the 
$e^+e^-\to \eta\gamma(\gamma)$ background, the condition  
$2E_{\gamma,\rm{max}}/\sqrt{s}<0.78$ is applied.
\begin{figure}
\centering
\includegraphics[width=0.47\textwidth]{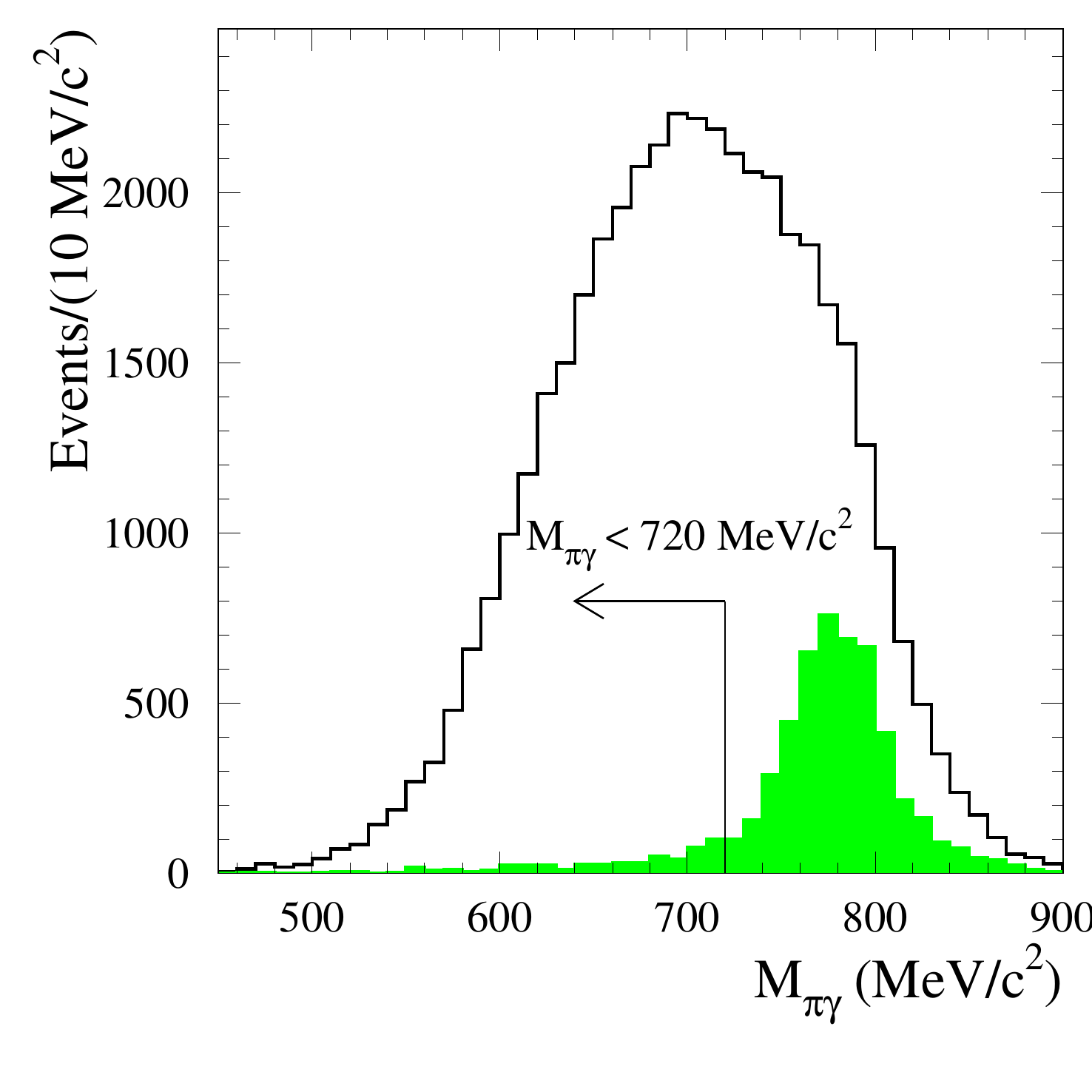}
\caption{ \label{fig4} The distribution of the invariant mass of the $\omega$
meson candidate for simulated events of the processes 
$e^+e^- \to \eta\pi^0\pi^0$ (open histogram)
and $e^+e^-\to \omega\pi^0(\gamma)$ (shaded histogram). 
The arrow indicates the selection criterion used.}
\end{figure}

Most of the $e^+e^-\to\omega\pi^0(\gamma)$ events remaining after applying
the condition $\chi^2_{\pi\pi\gamma}>60$ contain an additional photon 
emitted from the initial state at a large angle. To suppress this background,
the requirement of the absence of a $\omega$-meson candidate in an event 
is used. The $\omega$ candidate is defined as a combination of three photons,
one of which is the most energetic photon in the event, and the other two must
have an invariant mass in the range $|M_{2\gamma}-m_{\pi^0}|<35$ MeV. If there
are several $\omega$ candidates in an event, 
one with the lowest $|M_{3\gamma}-m_{\omega}|$ value is chosen.
The distribution of the invariant
mass of the $\omega$ candidate ($M_{\pi\gamma}$) for simulated 
signal and background $e^+e^-\to\omega\pi^0(\gamma)$ events is shown in 
Fig.~\ref{fig4}. Events, for which $M_{\pi\gamma}>720$ MeV, are rejected. 
This condition also suppresses the background from the process 
$e^+e^-\to\omega\pi^0\pi^0$ by a factor of about 10.

\begin{table}
\footnotesize
\caption{\label{tab1} The effect of the selection criteria applied successively
on data events ($N_{\rm data}$), estimated background ($N_{\rm bkg}$),
and signal calculated for $\sigma(e^+e^-\to f_1)=50$ pb ($N_{\rm sig}$).} 
\begin{center}
\setlength{\arraycolsep}{15mm}
\begin{tabular}{lccc}
\hline
Selection conditions & $N_{\rm data}$ & $N_{\rm bkg}$ & $N_{\rm sig}$ \\
\hline 
$\chi^2_{\eta\pi\pi}<35$              & 90 & 86   & 4.5 \\
$+~\chi^2_{\pi^0\pi^0\gamma}>60$       & 13 & 15.5 & 3.6 \\
$+~2E_{\gamma,\rm{max}}/\sqrt{s}<0.78$ & 10 & 9.4  & 3.2 \\
$+~M_{\pi\gamma}<720$ MeV               &  2 & 1.1  & 1.8 \\
\hline
\end{tabular}
\end{center}
\end{table}
The effect of the selection criteria applied successively
on data events, estimated background and signal calculated 
for $\sigma(e^+e^-\to f_1)=50$ pb is demonstrated in Table~\ref{tab1}.

Finally, two events are selected in data. Their distribution over the 12 
energy points in comparison with the simulated background distribution 
is given in Table~\ref{tab2}.
\begin{table}
\footnotesize
\begin{center}
\caption{\label{tab2}
The center-of-mass energy ($\sqrt{s}$), integrated luminosity ($L$), number of
selected data events ($N$), number of background events ($N_{\rm bkg}$)
calculated using simulation with the statistical error.}
\begin{tabular}{cccccccc}
\hline
$\sqrt{s}$ (GeV) &  $L$ (nb$^{-1}$) & $N$ & $N_{\rm bkg}$ &$\sqrt{s}$ (GeV) &  $L$ (nb$^{-1}$) & $N$ & $N_{\rm bkg}$ \\
\hline 
 1.200 & 1185 &  0  &  $0.054\pm0.010$ & 1.300 & 2220 &  0  &  $0.108\pm0.021$ \\
 1.225 &  577 &  0  &  $0.028\pm0.005$ & 1.325 &  559 &  0  &  $0.037\pm0.007$ \\
 1.250 &  467 &  0  &  $0.025\pm0.005$ & 1.350 & 1945 &  0  &  $0.124\pm0.022$ \\
 1.275 &  516 &  0  &  $0.022\pm0.005$ & 1.360 &  826 &  0  &  $0.080\pm0.012$ \\
 1.280 &  740 &  0  &  $0.045\pm0.008$ & 1.375 &  612 &  0  &  $0.057\pm0.008$ \\          
 1.282 & 3451 &  2  &  $0.252\pm0.044$ & 1.400 & 2024 &  0  &  $0.275\pm0.032$ \\ 
\hline
\end{tabular}
\end{center}
\end{table}

\section{\boldmath Cross section for $e^+e^-\to f_1(1285)$ 
and $f_1(1285)\to e^+e^-$ branching fraction}
It can be seen from Table~\ref{tab2} that the two selected data events are
located at the energy point corresponding to the maximum of the $f_1(1282)$
resonance, where the calculated background is 0.25 events.
The simulation reproduces background
contribution reasonably well. After applying the condition 
$\chi^2_{\eta\pi\pi}<35$, the number of selected data events coincides with
the number of background events estimated from simulation within about 10\% 
statistical uncertainty. The next three conditions in Table~\ref{tab1} suppress
background by a factor of about 80. For the main background process 
$e^+e^- \to \omega \pi^0 (\gamma)$, these conditions remove well reconstructed 
events with five true photons and a spurious photon from the beam background, 
and six-photon events with an extra photon emitted from the initial state. 
Small remaining background contains improperly reconstructed events, for 
example, with a splitting photon from the $\omega$ decay, which fraction may 
be simulated incorrectly. To estimate quality of simulation of the 
background-suppression conditions, we study a sample of background events 
with $\chi^2_{\eta\pi\pi}>35$, which are well reconstructed in 
the $e^+e^-\to 6\gamma$ hypothesis ($\chi^2_{6\gamma}<30$). The number of 
such events selected in data (465) is in reasonable agreement with the number 
expected from simulation ($422\pm 2$). The selection criteria listed in the 
three last rows in Table~\ref{tab1} reduce the number of data events by a 
factor of $18.6\pm 3.6$ and the number of simulated background events by a 
factor of $20.2\pm 0.3$. From the simulation-to-data ratio $1.08\pm 0.21$, we 
estimate that the MC simulation reproduces the effect of the background 
suppression conditions with a systematic uncertainty of about 20\%. Taking 
into account 10\% uncertainty due to the $\chi^2_{\eta\pi\pi}<35$  condition 
discussed above, we conclude that the systematic uncertainty of the background 
prediction for the selection criteria listed in Table~\ref{tab1} is better 
than 30\%.

The distribution of data events listed in Table~\ref{tab2} is 
fitted with a sum of signal and background distributions:
\begin{equation}
N_i^{\rm th}=\varepsilon\sigma_{\rm vis}(\sqrt{s_i})L_i+
N_{{\rm bkg},i}R_{\rm bkg}
\label{eq3}
\end{equation}
where $\varepsilon$ is the detection efficiency for the process
$e^+e^-\to f_1(1285)$, $L_i$ is the integrated luminosity in the
point with energy $\sqrt{s_i}$,
$\sigma_{\rm vis}$ is the $e^+e^-\to f_1(1285)$ visible cross section,
$N_{{\rm bkg},i}$ is the background distribution given in Table ~\ref{tab2}.
The background scale factor $R$ is set to unity and varied during the 
fit within its 30\% systematic uncertainty.

The visible cross section is calculated as follows:
\begin{equation}
\sigma_{\rm vis}(\sqrt{s})=\int_{0}^{x_{\rm max}}
W(s,x)\sigma_{\rm B}(\sqrt{s(1-x)}) dx,
\label{eq1}
\end{equation}
where $W(s,x)$ is the so-called radiator function, 
which describes the probability density for emission of photons with
the total energy $x\sqrt{s}/2$ from the initial state~\cite{rad1}.
The Born cross section, $\sigma_{\rm B}$, is parametrized as follows:
\begin{equation}
\sigma_{\rm B}(\sqrt{s})=\sigma(e^+e^-\to f_1)
\frac{m^2_{f_1}\Gamma^2}{(s-m^2_{f_1})^2+m^2_{f_1}\Gamma^2}
\frac{m^3_{f_1}P(s)^3}{s^{3/2}P(m^2_{f_1})^3}.
\label{eq2}
\end{equation}
where the cross section at the resonance maximum
$\sigma(e^+e^-\to f_1)=(12\pi/m^2_{f_1})\* B(f_1\to e^+e^-)$. In Eq.~(\ref{eq2})
we assume that $f_1\to \eta\pi^0\pi^0$ decay  proceeds through the 
intermediate state $a_0(980)\pi^0$. Therefore, $P(s)$ is the $a_0(980)$ 
momentum. The radiative corrections reduce the visible cross section at the 
resonance maximum by 20\% compared with the Born cross section.

The detection efficiency for $e^+e^-\to \eta\pi^0\pi^0$ events with the 
$a_0^0(980)\pi^0$ intermediate state depends on the energy and
decreases from 5.3\% at 1.2~GeV to 4.9\% at 1.282~GeV, and then to
1.9\% at 1.4~GeV. The steep decrease of the effeciency above the resonance
maximum is due to the decrease of the $e^+e^-\to f_1$ Born cross section.
At 1.4~GeV about 80\% of signal events events contain an energetic photon
emitted from the initial state (see Eq.~(\ref{eq1})).

For the $f_0(500)\eta$  mechanism, the efficiency is 25\% lower. Assuming that
$f_1 (1285)$ decay to the $\eta\pi^0\pi^0$ final state proceeds through these
two mechanisms, and the fraction of $a_0^0(980)\pi^0$ is $(73\pm 8)\%$, we 
obtain that the detection efficiency at $\sqrt{s}=1.282$~GeV is equal to 
$(4.6\pm 0.3)\%$. The quoted
model error is estimated as a difference between the efficiencies calculated
in the models $a_0^0(980)\pi^0$ and $a_0^0(980)\pi^0 + f_0(500)\eta$. 
A detailed study of the systematic uncertainty associated with the selection
of multiphoton events based on the kinematic fit was performed in
Ref.~\cite{SNDomegapi} using $e^+e^-\to \pi^0\pi^0\gamma$ events. 
Basing on this study, we estimate
that the systematic uncertainty on the detection efficiency due to inaccuracy
in simulation of the detector response does not exceed 5\%. Taking into 
account the branching fraction $B(f_1(1285)\to \eta\pi^0\pi^0)=(17.3\pm0.7)\%$,
the detection efficiency in Eq.~(\ref{eq3}) at $\sqrt{s}=1.282$~GeV is 
$\varepsilon=(0.79\pm0.08)\%$,
where the error includes all the uncertainties discussed above.

As a result of the fit to the distribution of data events listed in 
Table~\ref{tab2}, the following value of the cross section at the resonance 
maximum is obtained
\begin{equation}
\sigma(e^+e^-\to f_1)=45^{+33}_{-24}\mbox{ pb},
\label{result}
\end{equation} 
which is in agreement with the theoretical prediction 
$(31\pm 16)$~pb~\cite{rudenko2}.
The significance of the $e^+e^-\to f_1(1285)$ signal estimated by comparing
the log-likelihood values for the fits with and without the resonance is
found to be $2.5\sigma$. We consider our result as a first indication of the 
process $e^+e^-\to f_1(1285)$.

The obtained value of $\sigma(e^+e^-\to f_1)$ corresponds to
the branching fraction
\begin{equation}
B(f_1(1285)\to e^+e^-)=(5.1^{+3.7}_{-2.7})\times 10^{-9}. 
\label{result1}
\end{equation} 
Since the significance of the $f_1(1285)$ signal is not large, we also quote
the 90\% confidence level upper limit
\begin{equation}
B(f_1(1285)\to e^+e^-) < 9.4\times 10^{-9}.
\label{result2}
\end{equation}

\section{Summary}
The search for the direct production of the $f_1(1285)$ resonance in 
$e^+e^-$ collisions is performed using the data sample with an
integrated luminosity of 15.1 pb$^{-1}$ recorded in the SND experiment
at the VEPP-2000 $e^+e^-$ collider in the energy region 
$\sqrt{s}=1.2$--1.4 GeV. About 3.5 pb$^{-1}$ were collected
at the maximum of the $f_1(1285)$ resonance.
To search for the process $e^+e^-\to f_1(1285)$, the decay mode
$f_1(1285)\to\eta\pi^0\pi^0$ has been used. After applying the selection
criteria, two events have been observed at the peak of the $f_1(1285)$
resonance and zero events beyond the resonance. These two events correspond
to the cross section $\sigma(e^+e^-\to f_1)=45^{+33}_{-24}$ pb and the 
branching fraction $B(f_1(1285)\to e^+e^-)=(5.1^{+3.7}_{-2.7})\times 10^{-9}$.
The significance of the $e^+e^-\to f_1(1285)$ signal is $2.5\sigma$.
We consider this result as a first indication of the process 
$e^+e^-\to f_1(1285)$. The measured branching fraction agrees with the 
theoretical prediction~\cite{rudenko2}.

\end{document}